\documentclass[aps,pra,twocolumn,reprint]{revtex4-1}

\usepackage{graphicx}   
\usepackage{color}      
\usepackage{subfigure}  
\usepackage{hyperref}   
\usepackage{tikz}
\hypersetup{
    colorlinks,
    linkcolor={red!50!black},
    citecolor={blue!50!black},
    urlcolor={blue!80!black}
}
\usepackage{booktabs}
\usepackage{tikz} 
\usepackage{amsmath}
\usepackage[utf8]{inputenc}
\usepackage{amssymb}
\usepackage{dsfont}
\usepackage{siunitx}
\usepackage{import}

\usepackage{braket}
\usepackage{mathcomp}

\newcommand{\te}[1]{\text{#1}} 
\newcommand{\rr}{$\Rightarrow$}
\newcommand{\bdef}[2]{\underbrace{#1}_{\equiv #2}}
\newcommand{\ke}{\ket{1}}
\newcommand{\kz}{\ket{2}}
\newcommand{\kd}{\ket{3}}
\newcommand{\kv}{\ket{4}}
\newcommand{\forcenewpage}{\textcolor{white}{1}\\
\newpage }
\newcommand{\SF}[1]{\framebox[1.04\width][l]{\textbf{#1}}}
\newcommand{\rot}[1]{\mathbf{\nabla}\times #1} 
\newcommand{\di}[1]{\mathbf{\nabla}\cdot #1} 
\newcommand{\grad}[1]{\mathbf{\nabla} \left( #1 \right)} 
\newcommand{\nab}{\vec{\nabla}} 
\newcommand{\ddd}[1]{\frac{\partial^2}{\partial {#1}^2}}
\newcommand{\vx}{\mathbf{x}}
\newcommand{\che}{\chi^{(1)}}

\newcommand{\lslash}{\lambdabar}

\renewcommand\vec[1]{\mathbf{#1}}
\renewcommand{\Re}[1]{\text{Re} [ #1 ] }
\renewcommand{\Im}[1]{\text{Im} [ #1 ] }
\newcommand{\xo}{\chi^{(1)}}
\newcommand{\LV}{\mathcal{L}}
\newcommand{\Om}{\Omega}
\newcommand{\rs}{\hat{\rho}_a}
\newcommand{\trace}[2]{Tr_{#1} \lbrace #2 \rbrace}

\graphicspath{{pics/}{plots/}}
 
\begin{document}
\title{Optimal pulse propagation in an inhomogeneously gas-filled hollow-core fiber 
}

\today

\author{Roman Sulzbach}
\email[Corresponding author:]{sulzbach@gfz-potsdam.de}
\altaffiliation{Now at:\\ Deutsches Geoforschungszentrum (GFZ), Telegrafenberg,\\ Potsdam D-14473, Germany}
\affiliation{Institut für Angewandte Physik, TU Darmstadt, Hochschulstraße 4A, Darmstadt D-64289, Germany}
\author{Thorsten Peters}
\affiliation{Institut für Angewandte Physik, TU Darmstadt, Hochschulstraße 4A, Darmstadt D-64289, Germany}
\author{Reinhold Walser} 
\affiliation{Institut für Angewandte Physik, TU Darmstadt, Hochschulstraße 4A, Darmstadt D-64289, Germany}
\begin{abstract}
We study optical pulse propagation through a 
hollow-core fiber filled with a radially inhomogeneous cloud of cold atoms. A co-propagating control field 
establishes electromagnetically induced transparency.
In analogy to a graded index fiber, the pulse experiences micro-lensing and the transmission spectrum becomes distorted.
Based on a two-layer model of the complex index of refraction, we can 
analytically understand the cause of the aberration, which is 
corroborated by numerical simulations for a radial Gaussian-shaped 
function. With these insights, we show that the spectral distortions 
can be rectified by choosing an optimal detuning from one-photon 
resonance.
\end{abstract}
\maketitle
\section{Introduction}
Tight transverse confinement of atoms and light fields over macroscopic distances produces strong light-matter coupling. 
In recent years, this has been achieved by 
loading laser-cooled atomic ensembles into hollow-core fibers (HCFs) 
\cite{RMV95,RZD97,MCA00,CWS08,VMW10,BHP11,PBH12,BHP14,OTB14,BSH16,LWK18,HPB18,NLW18,XLC18,YB19pre,HPL19pre} . This strong coupling can then be exploited to observe nonlinear optical effects at the few-photon level \cite{CVL14,HCG11,HCG12},
 or to create strongly correlated photonic quantum gases \cite{CGM08,KH10b,AHK11,HA12,HNR12,AHC13} to mention a few.

The coupling strength between a resonant light field and an atom 
depends on the ratio of the light-mode cross-section 
and the atomic absorption cross-section. The closer the light-mode 
matches the atomic cross-section the more likely it is that a photon is absorbed by an atom \cite{CVL14}. Therefore, the focus lies in 
recent years on using either small-core photonic bandgap fibers with 
core diameters $d\sim \SI{10}{\micro \meter}$ 
\cite{CWS08,VMW10,BHP11,PBH12,BHP14,BSH16,YB19pre}, or medium-core 
fibers with core diameters of several \SI{10}{\micro \meter} 
\cite{OTB14,LWK18,HPB18,NLW18,XLC18,HPL19pre}, instead of large-core 
capillaries. In order to prevent collisions of the laser-cooled atoms with the fiber wall at room temperature, the atoms are usually guided into the fiber by a Gaussian-shaped, red-detuned, far-off resonant optical trap (FORT). As the light-mode diameter of the FORT is in the range of the fiber core diameter and the temperature of the atomic 
ensemble inside the HCFs is usually much smaller than the trap depth, 
the atomic density distribution is strongly radially dependent across the light-mode cross-section. For a thermalized atomic ensemble, 
e.\,g., a Gaussian radial density distribution can be expected \cite{GWO00}. This, in principle, requires to consider the radially varying index of refraction when calculating the light propagation, as frequency-dependent lensing can occur \cite{VP05,RHK15}. For a purely absorptive medium this has been done recently \cite{GRR18}.

\begin{figure}[t]
\def\svgwidth{\columnwidth}
\includegraphics[scale=1.]{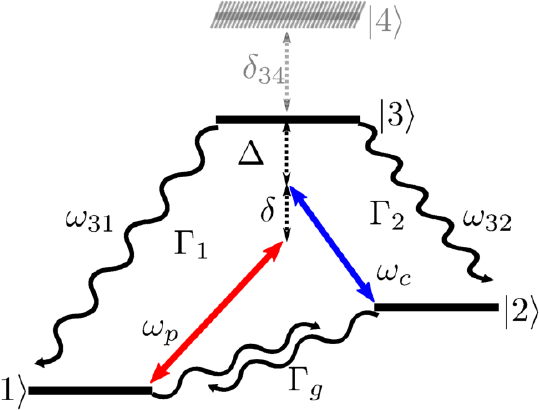}
\caption{
\label{fig:CS}
Atomic level scheme in Raman configuration with transition frequencies $\omega_{31}=\omega_3-\omega_1$ and $\omega_{32}=\omega_3-\omega_2$ interacting with  the control pulse $\Omega_c$ and probe pulse $\Omega_p$. The one- and two-photon detunings are denoted by $\Delta$ and $\delta$, respectively.
Spontaneous decay rates $\Gamma_1$ and $\Gamma_2$ couple the excited state $\ket{3}$ to the ground states $\ket{1}$ and $\ket{2}$.
Residual perturbations are accounted for by a ground-state decoherence rate $\Gamma_g$. 
Experimentally, the presence of level $\ket{4}$ can be relevant, but will not be considered at present.}
\end{figure}
All the aforementioned proposals using HCFs \cite{HCG11,HCG12,CGM08,KH10b,AHK11,HA12,HNR12,AHC13} rely on establishing electromagnetically induced transparency (EIT) \cite{H97,FIM05} within the atomic ensemble as to allow for strong photonic nonlinearities while suppressing linear absorption. The general $\Lambda$-type level structure for EIT is shown in Fig.~\ref{fig:CS}. 
The two metastable ground states $\ket{j=1,2}$ are coupled to the excited state $\ket{3}$ by a weak probe field, denoted by its Rabi frequency $\Omega_p(\omega,r)$, and a strong control field, denoted by its Rabi frequency $\Omega_c(r)$, respectively, as defined in App.~\ref{App:IAP}. Note that $\Omega_p(\omega,r)$ depends on the frequency of the probe pulse as well as on the radial position.
The strong control field modulates the refractive index for the weak probe field on transition $\ket 1 \leftrightarrow \ket 3$. As this modulation is dependent on the control Rabi frequency, a spatially varying control beam results in a spatially modulated refractive index. Incoherent interaction with the vacuum field leads to decay from the excited state $\ket{3}$ to the ground states $\ket{j=1,2}$
with rates $\Gamma_{j}$. When no ground state decoherence or dephasing processes are present EIT leads to perfect transmission at exactly the two-photon resonance between states $\ket{1}$ and $\ket{2}$. The effect of a radially varying control beam intensity in a homogeneous ensemble on the radial propagation dynamics of a probe beam has been studied both experimentally and theoretically \cite{MSF95,JMK95,MSF96,TFR99,VP05,SRZ06,PUG08b,VSH09,SLR10,DE11,ZDY11,SWF05}. It was shown that a spatial variation of the control beam intensity in EIT can be used, e.g., to achieve electromagnetically induced focusing \cite{MSF95,MSF96}.

In this work we analytically and numerically study the propagation of a weak light field under EIT conditions with a Gaussian-shaped control beam intensity and a radially dependent atomic density distribution of comparable width. To the best of our knowledge, a combination of spatially-dependent control Rabi frequency as well as atomic density distribution has not yet been considered analytically. However, in a recent work micro-lensing under EIT conditions has been observed in a HCF and studied numerically \cite{NLW18}. Although our work is applicable not only to cold atomic ensembles loaded into HCFs, we specifically discuss the results in the context EIT in HCFs. The reason is that so far, narrow-band EIT (using laser-cooled atoms) and related effects have been observed in HCFs by several groups \cite{BHB09,BSH16,LNK17,NLW18}, but the experimental results have not yet been compared to a thorough theoretical analysis including the radial propagation dynamics. As we will show, the spatially varying index of refraction due to the inhomogeneous atomic density and the control Rabi frequency can lead to frequency-dependent lensing and distortion of the probe beam. However, by a correct choice of parameters, these effects can be mitigated.

This paper is organized as follows: Sec.~\ref{sec:exp_setup}
briefly discusses the considered experimental setup. By estimating the order of magnitude of its physical properties, we thereby establish the conditions for the theoretical model. In Sec.~\ref{sec:Bloch}, we formulate 
 the EIT response of the atomic medium to the light fields and  derive a propagation equation in Sec.~\ref{sec:parax}. 
In Sec.~\ref{sec:lprop}, we use these results to quantify and mitigate the lensing effects in an atom-filled HCF in the EIT regime. The results are summarized in Sec.~\ref{sec:sum}.

\section{Considered experimental setup}~\label{sec:exp_setup}


Experimentally, HCFs are loaded with
laser-cooled atoms from a magneto-optical trap 
\cite{CWS08,VMW10,BHP11,PBH12,BHP14,OTB14,BSH16,YB19pre,LWK18,HPB18,NLW18,XLC18,HPL19pre}. 
The HCFs have an inner radius in the range of
 $\SI{3.5}{\micro \meter}$ 
$\lesssim a_0 \lesssim \SI{25}{\micro\meter}$
 and a length $L$ in the range of centimeters. For wavelengths
 $\lambda_p \sim \SI{1}{\micro \meter}$, the fibers have a nearly Gaussian eigenmode that we will refer to as $u_e^{(1)}(r)$. In the following, we will establish a simple model for the single-mode HCF assuming its walls to be ideally conducting, while we externally suppress propagation of higher fiber modes. Due to the small radial dimension of the fiber, we introduce the reduced wavelength $\lslash_p =\lambda_p/2 \pi= k_p^{-1}$ and define a fiber parameter \cite{OWT83} as 
\begin{equation}
\label{eq:fibpar}
v = a_0 k_p.
\end{equation}
It compares the wavelength to the fiber radius, which will have a critical influence on light propagation as we will show later. In the current experimental cases, we have $v\sim 25$ \cite{BHB09,BHP11,BHP14,BSH16,YB19pre} to $v\sim200$ \cite{LNK17,HPB18,NLW18,HPL19pre}.

In order to avoid collisions of the cold atoms with the fiber-core wall at room temperature, a FORT is created inside the HCF by a red-detuned laser field and a mode field radius $\sigma_d$. With the available laser power a trap depth of $T_\text{FORT}$ of up to several mK is achieved. Because the induced dipole potential depth 
$V_0=k_B T_\text{FORT}$ is deep compared to the kinetic energy of the atoms, the resulting thermalized atomic density distribution
\begin{align}
\label{eq:densdis}
n_a(r) &= n_0 e^{ - \frac{r^2}{\sigma_a^2} },
\end{align}
has a radial Gaussian shape varying with the distance $r$ from the fiber axis and a width $ 
\sigma_a(T) = \sigma_d \sqrt{T /T_\text{FORT}}$ depending on the potential and the temperature of the atoms inside the HCF (see App.~\ref{ap:atomicdistr}). Integrating the density over the volume of the fiber yields the total 
particle number 
$N_a=\pi \sigma_a^2 L n_0. $ For around 
$N_a=\num{2.5e5}$ atoms loaded into the HCF and an atomic temperature of several hundred $\mu$K \cite{BHP14}, this corresponds to an atomic peak density of around $
n_0 \sim \SI{e12}{\centi \meter\tothe{-3}}$.

\begin{figure}
\def\svgwidth{\columnwidth}
\includegraphics[scale=1]{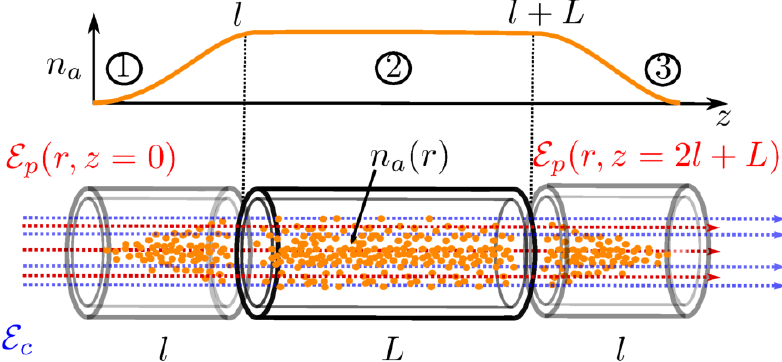}
\caption{
 Propagation of a probe pulse $\mathcal{E}_p$ (red) through a HCF of length $2l+L$ with core radius $a_0$. The fiber is filled with an inhomogeneous cold gas of density $n_a(r,z)$. Along the propagation direction, the fiber is divided
into the entrance and exit zones \textcircled{1}, \textcircled{3}, where the atomic density rises slowly to its homogeneous value $n_a(r)$ in \textcircled{2}.  The polarizability is controlled by the copropagating immutable control beam $\mathcal{E}_c$ (blue). The inhomogeneous, frequency dependent susceptibility will alter the mode shape and phase of the probe pulse.}
 \label{fig:system}
\end{figure}
Throughout this paper we will assume that the atomic density $n_a(r)$ is basically independent of the longitudinal position $z$ inside the HCF, apart from the short regions $l\ll L$ near the HCF entrance and exit (see Fig.~\ref{fig:system}). In these regions the density decreases quickly, but still adiabatically, towards zero. This, for instance, can be due to the quickly decreasing dipole potential outside the HCF from the diverging dipole trap beam, the inhomogeneous density above the fiber, or due to optical pumping or pushing away of the atoms left outside of the fiber.

Once the atoms are loaded into the fiber, the FORT is switched off rapidly before the start of any measurements to avoid
ac-Stark shifts by the strong FORT. This leads to a ballistic expansion and loss of the atoms due to collisions with the fiber wall on a timescale of $\tau_\text{loss}\sim a_0/\sqrt{2k_BT/m_\text{atom}}$. If the timescale of the measurements is sufficiently short 
$t< \tau_\text{loss}$, then one can assume the density $n_a(r)$ is time-independent during the measurement.

\section{Electromagnetically induced transparency\label{sec:Bloch}}

In our EIT experiment, the control field is given by
\begin{equation}
\mathbf{E}_c(\vx, t)=
\Re{
\mathbf{e}_c  
e^{-i (\omega_c t - k_c z)} u^{(1)}_e(r)\mathcal{E}_c},
\end{equation}
and co-propagates with the probe field along the $z-$direction.  
It is linearly polarized along $\mathbf{e_c}$, monochromatic with frequency $\omega_c$, and has a complex amplitude 
$\mathcal{E}_c$. Due to the cylindrical symmetry of the fiber, the field depends only on $(z,r=\sqrt{x^2+y^2})$. Unaffected by the atomic medium, this field propagates in a mode $u^{(1)}_e(r)$, as in an empty fiber.
Assuming EIT conditions, the atomic gas becomes then transparent and ultra-dispersive for the weak probe pulse
\begin{equation}
\label{eq:Epin}
\mathbf{E}_p^{(in)}(x,y,t) = \Re{
  \mathbf{e}_p     
e^{-i\omega_p t - \frac{t^2}{\tau_p^2}}u^{(1)}_e(r)\mathcal{E}_p},
\end{equation}
specified 
at the entrance side of the fiber at $z=0$ with $|\mathcal{E}_p| \ll |\mathcal{E}_c|$. 
Here, the  probe pulse lasts for a duration $\tau_p$, has a linear 
polarization $\mathbf{e}_p$ and a carrier frequency $\omega_p$.

As EIT is a highly frequency sensitive effect, we will study the propagation of a probe pulse 
$\mathbf{E}_p(\vx, \omega)$ in the frequency domain. 
However, when the pulse emerges from the fiber, one  observes the temporally delayed, real time signal 
$\mathbf{E}_p(\vx, t)$. Clearly, both fields are connected through the temporal Fourier transformation
\begin{align}
\mathbf{E}_p(\vx, t) &= \int_{- \infty}^\infty d \omega 
  \frac{e^{- i \omega t}}{\sqrt{2 \pi}} \mathbf{E}_p(\vx, \omega).~\label{eq:FourOm}
\end{align}
As all physical fields are real-valued, we have the auxiliary constraint that 
$\mathbf{E}_p(\vx, -\omega)=\mathbf{E}_p^\ast(\vx, \omega)$.
Inside the HCF, the confined atomic gas exhibits a radial density variation. Thus, we describe the wave propagation in the spatial domain. 
In this sense, we find that the boundary condition for the probe pulse at the fiber entrance of Eq.~\eqref{eq:Epin} reads in the frequency domain
\begin{align}
\label{Epfourier}
\mathbf{E}_p^{(in)}(x,y,\omega) &= 
\mathbf{e}_p  u^{(1)}_e(r)\left[
\mathcal{E}_p^{(in)}(\omega)+
\mathcal{E}_p^{(in)\ast}(-\omega)\right] ,\\  
\mathcal{E}_p^{(in)}(\omega)&=
 \frac{e^{-\left(\frac{\omega-\omega_p}{\Delta\omega}\right)^2}}{ \sqrt{2}\Delta\omega} \mathcal{E}_p.
\end{align}
These are narrow Gaussian envelopes at $\omega=\pm\omega_p$ with a frequency width $\Delta\omega=2/\tau_p\ll\omega_p$.

Due to the low temperature of the atoms within the HCF, we can neglect their spatial displacement during the probe pulse duration.
Thus, the atomic gas responds locally to an applied  probe field $\mathbf{E}_p(\vx,\omega)$. All spatial variations are accounted for by the corresponding amplitude change. Consequently, we will suppress the position parameter in the following. 

For weak probe pulses the atomic gas responds linearly with a polarization density
\begin{align}
\label{chi1}
\mathbf{P}_p(\omega)
 = \epsilon_0 \ \che(\omega)
  \mathbf{E}_p(\omega),
\end{align}
defining a linear complex susceptibility
$\che(\omega)$ on the p-branch using the vacuum permittivity $\epsilon_0$. 
Microscopically, the polarization density 
\begin{equation}
\mathbf{P}_p(\omega) = n_a\mathbf{d}(\omega)
\end{equation}
is obtained from a small local sample of atoms, weighing the particle number distribution $n_a$ with the average
atomic dipole moment
$\mathbf{d}=
\text{Tr}\{\hat{\mathbf{d}} \hat{\rho}\}$. Here, we specify the state of the atomic ensemble with the single particle density operator 
$\hat{\rho}$.
Eventually, this defines the optical susceptibility tensor on the probe transition as
\begin{equation}
\label{eq:polarizability}
\che_{rs} =  
\frac{n_a}{\epsilon_0}
\frac{\partial d_r}{\partial E_{p,s}}|_{E_p=0},
\end{equation}
with respect to the Cartesian coordinates $r,s$.

\subsection{Optical master equation}
The dynamics of the Raman transition depicted in Fig.~\ref{fig:CS} are given by the interplay of the 
coherent probe $\Omega_{p}$ and control pulses $\Omega_{c}$ with dissipation. The Rabi frequencies here measure the effective dipole coupling strength of the transition at the pump and probe frequencies 
$\omega_p$ and $\omega_c$, respectively. Within the rotating-wave approximation,
one obtains the Hamiltonian matrix 
\begin{equation}
\label{eq:HCS}
H' =  \hbar \  
\begin{pmatrix}
\Delta + \delta &  0 &\frac{\Omega_{p}^*}{2}\\
0 &  \Delta & \frac{\Omega_{c}^*}{2}\\
\frac{\Omega_p}{2} & \frac{\Omega_{c}}{2} & 0
\end{pmatrix},
\end{equation} 
in a suitable interaction frame (cf. App.~\ref{App:IAP}), introducing the two-photon detuning $\delta = \omega - \omega_c - \omega_{21}$ and the 
one-photon detuning $\Delta = \omega_{c} - \omega_{32}$, where $\omega_{ij}$ is the transition frequency between levels $\ket{i}$ and $\ket{j}$. The condition of two-photon resonance 
$\omega_p=\omega_c+\omega_{21}$, 
defines the carrier frequency of the probe beam. 

The basic mechanism of EIT \cite{QUOPSCUL,FIM05,cohentannoudjiBOOK1} follows from the three dressed eigenstates  
$\{\ket{D},\ket{+}, \ket{-}\}$ of 
$H'$ 
at two-photon resonance $\delta=0$. 
In particular, one finds the dark state
$\ket{D} = \cos{\theta} \ket{1} - \sin\theta \ket{2}$, as a superposition of ground states mixed at an angle 
$\tan \theta = \Omega_p/\Omega_c$. Preparing a system in this state implies that there are
no allowed dipole transitions to other states of the manifold as $\bra{\pm} \hat{\vec{d}} \cdot\vec{E} \ket{D}=0$ and  that 
$\ket{D}$ is immune to spontaneous emission from the excited state
 $\ket{3}$. 

Embedding atoms in an open environment, 
introduces fundamental, as well as technical decoherence \cite{cohentannoudjiBOOK1,sturm14,rosenbluh98,walser294,nandi604}. Thus, one needs to use a master equation for the density operator $\hat{\rho}$

\begin{align}
\dot{\hat{\rho}} =& 
-\frac{i}{\hbar} [H', \hat{\rho}]+ 
\sum_{i=1}^2 
\Gamma_{i} (\hat{\sigma}_{i3} \hat{\rho} \hat{\sigma}_{i3}^\dagger - \tfrac{1}{2}\hat{\sigma}_{i3}^\dagger \hat{\sigma}_{i3} \hat{\rho} - \tfrac{1}{2}\hat{\rho}  \hat{\sigma}_{i3}^\dagger \hat{\sigma}_{i3})\notag\\
\label{mastereq}
&+ 
\Gamma_{g} \left( \hat{\sigma}_{ii} \hat{\rho} 
\hat{\sigma}_{ii}- \tfrac{1}{2}
 \hat{\sigma}_{ii} \hat{\rho} 
- \tfrac{1}{2}\hat{\rho}  \hat{\sigma}_{ii}\right).
\end{align}
While the first term describes the coherent dynamics, the second term represents 
spontaneous emission from the excited state to the ground states with rates $\Gamma_{1}$ and $\Gamma_{2}$, respectively. Stimulated processes are not relevant at room temperature. The third term models the experimentally relevant ground state dephasing occurring at 
rate $\Gamma_g$ due to transit-time broadening \cite{BSH16} or decoherence due to magnetic field gradients.  This breaks the stationarity of the dark state leading to finite absorption under EIT conditions.
By regrouping the atomic density matrix elements as a column vector $\boldsymbol{\varrho}=(\rho_{11}, \rho_{12}, \rho_{13},
\rho_{21}, \rho_{22}, \rho_{23}, \rho_{31}, \rho_{32}, 
\rho_{33})$, one obtains 
\begin{align}
\label{eq:mast}
\partial_t\boldsymbol{\varrho}=& i L \boldsymbol{\varrho},
\end{align}
with a system Liouville matrix $L\in \mathds{C}^{9\times9}$ that reads 
\begin{widetext}
\begin{small}
\begin{align}
\label{eq:mast2}
L&=\left(
\begin{array}{ccccccccc}
 0 & 0 & \frac{\Omega _p}{2} & 0 & 0 & 0 
 & -\frac{\Omega _p^*}{2} & 0
   & -i \Gamma _1 \\
 0 & i \Gamma _g-\delta  & \frac{\Omega _c}{2} & 0 & 0 & 0 & 0 &
   -\frac{\Omega _p^*}{2} & 0 \\
 \frac{\Omega _p^*}{2} & \frac{\Omega _c^*}{2} &
   i \frac{\Gamma}{2}-\delta -\Delta & 0 & 0 & 0 & 0 & 0 
   & -\frac{\Omega_p^*}{2} \\
 0 & 0 & 0 & \delta + i \Gamma _g & 0 
 & \frac{\Omega _p}{2} 
 & -\frac{\Omega_c^*}{2} & 0 & 0 \\
 0 & 0 & 0 & 0 & 0 & \frac{\Omega _c}{2} & 0 
 & -\frac{\Omega _c^*}{2}
   & -i \Gamma_2 \\
 0 & 0 & 0 & \frac{\Omega _p^*}{2} 
 & \frac{\Omega_c^*}{2} & i \frac{\Gamma}{2}-\Delta  & 0 & 0 
 & -\frac{\Omega_c^*}{2} \\
 -\frac{\Omega _p}{2} & 0 & 0 & -\frac{\Omega _c}{2} & 0 & 0 & \delta +\Delta +
 i  \frac{\Gamma}{2}  & 0 & \frac{\Omega _p}{2} \\
 0 & -\frac{\Omega _p}{2} & 0 & 0 & -\frac{\Omega _c}{2} & 0 & 0 & \Delta +
 i \frac{\Gamma}{2} & \frac{\Omega _c}{2} \\
 0 & 0 & -\frac{\Omega _p}{2} & 0 & 0 & -\frac{\Omega _c}{2} 
 & \frac{\Omega_p^*}{2} & \frac{\Omega _c^*}{2} 
   & i \left(\Gamma
   _1+\Gamma _2\right) \\
\end{array}
\right).
\end{align}
\end{small}
\end{widetext}
Here, we have defined $\Gamma=\Gamma_1 +\Gamma_2+\Gamma_g$. For simplicity,  we will assume in the following an equal branching ratio 
$\Gamma_1 = \Gamma_2=\Gamma_0/2 $, which defines a
time scale $\tau_0=1/\Gamma_0$ for the irreversible atomic relaxation of state $\ket{3}$. 

\subsection{Linear susceptibility}
\begin{figure}[t]
\includegraphics[scale=1]{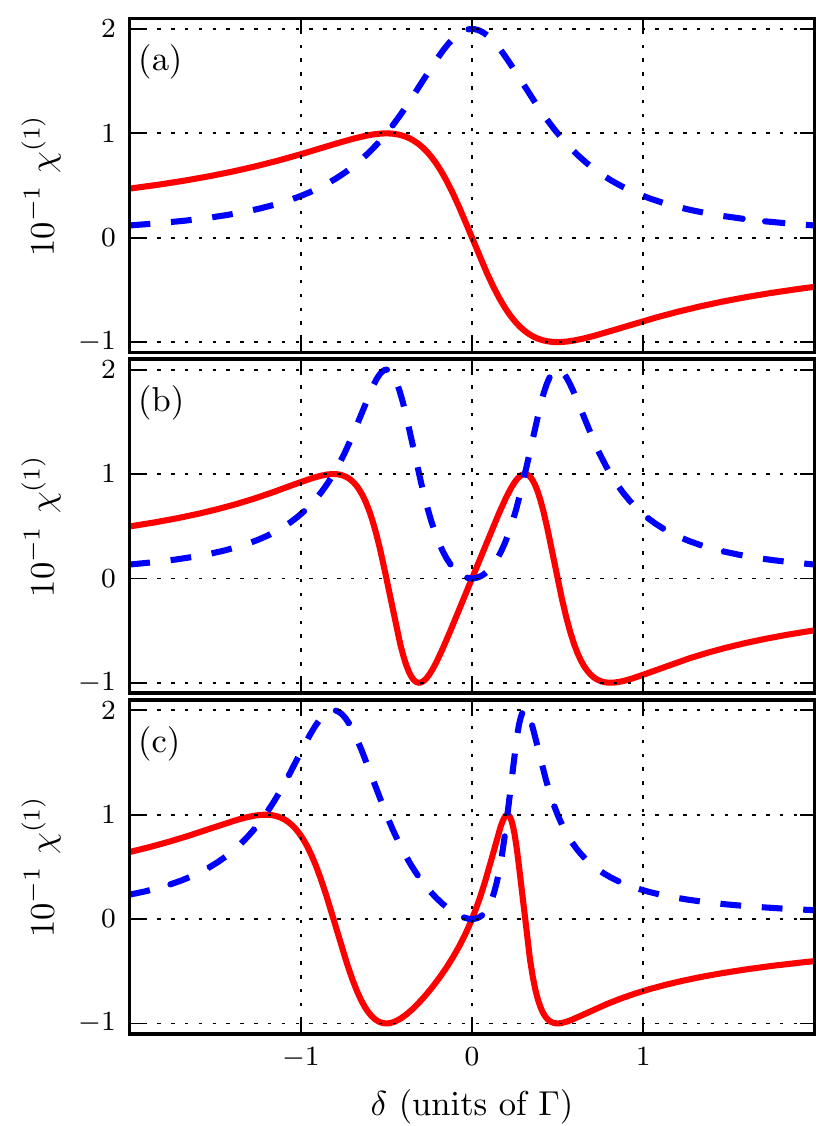}
\caption{
\label{fig:che}
Real part $\chi'(\delta)$ (solid red) and imaginary part $\chi''(\delta)$ (dashed blue) of the complex 
susceptibility versus two-photon detuning $\delta$ for: (a) ($\Omega_c,\Delta$, $\Gamma_g$) = $(0, 0, 0)\Gamma$,  
(b) ($\Omega_c,\Delta$, $\Gamma_g$) = $(1, 0, 0)\Gamma$,
(c) $(\Omega_c,\Delta, \Gamma_g) = (1, 1, 0)\Gamma$, with 
$\che_0 = \num{0.1}$.
\label{fig:polariz}}
\end{figure}
The Rabi frequency $\Omega_p$ 
arises from the weak probe pulse of duration $\tau_p$ propagating through the HCF. Therefore, there is also a slow implicit time dependence present. 
However, we want to consider pulses lasting longer than the atomic 
relaxation time  $\tau_p\gg \tau_0$. 
In this case, the atomic system is in equilibrium 
$\boldsymbol{\varrho}^s$ with respect to the instantaneous field 
\begin{equation}~\label{eq:steady_state}
L\boldsymbol{\varrho}^s=0,
\end{equation}
and the system remains in steady state when parameters change slowly.
From the steady state solution of Eq.~\eqref{eq:steady_state}, one obtains the polarization density as
\begin{align}
\label{pol}
\vec{P}_p(\omega)=n_a \vec{d}_{13} \rho_{31}^{s}.
\end{align}
From now on we will refer synonymously to the frequency of the 
pulse either by $\omega$, or through the two-photon detuning 
$\delta = \omega - \omega_p$.
In the limit of weak probe fields  $|\Omega_p| \ll \Gamma$ the 
linear susceptibility from Eq.~\eqref{chi1} reads
\begin{align}
\label{eq:chedef}
\che& =\che_0 f, 
&\che_0 = n_a \alpha_0,
&& \alpha_0=\frac{|d_{13}|^2 S_{FF'}}{2 \epsilon_0 \hbar \Gamma}, 
\end{align}
where the dimensionless shape function
\begin{equation}
\label{eq:F}
f(\delta, \Delta,|\Omega_c|)= 
\frac{(\delta+ i \Gamma_g ) \Gamma}{
\frac{|\Omega_c|^2}{4}- (\delta+ i \Gamma_g )((\delta + \Delta)+i\frac{\Gamma}{2})}
\end{equation}
contains all frequency dependencies. Please note that we have introduced \emph{ad hoc} in 
Eq.~\eqref{eq:chedef} the hyperfine transition strength factor $S_{FF'}$. It considers that our effective three-level system is formed by manifolds of hyperfine states.  
For negligible ground-state dephasing $\Gamma_g=0$, and close to two-photon resonance, this 
reduces to 
\begin{align}
\label{fvanishgammag}
f &=\delta \Gamma
\frac{
\frac{|\Omega_c^2|}{4}-\delta  (\delta +\Delta )
+i\delta \frac{\Gamma}{2}
}{
\left[\frac{|\Omega_c^2|}{4}-\delta  (\delta +\Delta )\right]^2+ \delta ^2 (\frac{\Gamma }{2})^2}\\
&=
\underline{\delta}\left(1+\frac{\Delta }{\Gamma} \underline{\delta}\right)
   +
   i
   \frac{\underline{\delta}^2}{2}\left(1+
   \frac{2 \Delta }{\Gamma}  \underline{\delta}\right)+\ldots \notag
\end{align}
where the detuning  $\underline{\delta} =\delta/\delta_\text{EIT}$  is specified in units  of the transparency window width
$\delta_\text{EIT}=|\Omega_c|^2/4 \Gamma$.
From the real part of Eq.~\eqref{fvanishgammag}, one can also 
determine the left and right zero crossings at 
$\delta^\pm=\pm(\sqrt{\Delta^2+|\Omega_c|^2}\mp\Delta)/2$.
Within those limits the absorption is a convex function .

In general, the linear susceptibility  $\che= \chi^\prime + i \chi^{\prime \prime}$, is a complex, analytic function
whose real part $\chi^\prime$ leads to refraction and whose imaginary part $\chi^{\prime\prime}$ causes absorption.
The magnitude of the susceptibility is of the order $\che_0 \sim 10^ {-2}\ll 1$.  
In Fig.~\ref{fig:polariz}, we depict the typical dependence of the susceptibility on the two-photon detuning $\delta$. 
Without the control beam, $\Omega_c=0$, this leads to the conventional
Lorentzian absorption and dispersion spectrum with a full width at half maximum (FWHM) spectral size $\Gamma$, shown in Fig.~\ref{fig:polariz}(a). In Fig.~\ref{fig:polariz}(b), 
we irradiate the gaseous sample with a control beam Rabi frequency $\Omega_c=\Gamma$. 
This opens a transparency window around $\delta=0$ 
and strong anomalous dispersion reduces of the speed of light drastically
 \cite{HHD99}. The spectrum becomes asymmetric 
when driving the system out of one-photon resonance as seen in
Fig.~\ref{fig:polariz}(c). A finite decay rate 
$\Gamma_g\neq 0$ would lead to residual absorption, even at $\delta=0$. 
\section{Light propagation}
\label{sec:parax}

A gas-filled HCF is analogous to an dielectric wave guide
with a graded index (GRIN) medium \cite{BornWolf99,OWT83}. 
The propagation equation for the electric field can be 
obtained from the macroscopic Maxwell equations of a nonmagnetic, but linear dielectric  material with electric permittivity 
$\varepsilon_r(\vx,\omega)=1+\che(\vx,\omega)$ from 
Eqs.~(\ref{chi1}) and (\ref{pol}).
The spatial dependence of the susceptibility arises on the one side from the spatial density variation $n_a(\vx)$ in Eqn.~\eqref{eq:densdis} and on the other side from the mode profile of the control field 
$\Omega_c(\vx)$. Considering these aspects of the setup, one finds a vectorial Helmholtz-type equation for the probe field
\begin{align}
\label{eq:W1}
\left( \nabla^2+k^2\varepsilon_r(\vx,\omega)  \right)  
\mathbf{E}_p(\vx, \omega) =
-\nabla (\mathbf{E}_p\cdot \nabla \log{\varepsilon_r} ), 
\end{align}
where  $k \equiv \omega/c_0$ describes the vacuum dispersion and  
$c_0$ the speed of light in vacuum.

The term on right hand side of the equation is important for stratified media \cite{BornWolf99} with a significant gradient of the permittivity in the propagation direction. 
However in a GRIN fiber, where the gradient is in the radial direction, this can be neglected as long as the wavelength $\lambda$ is much smaller than the characteristic scale of the longitudinal density variation or the radius of the fiber $a_0$.
This criterion is still satisfied for typical fiber parameters 
$v \gg 1$ from Eq.~\eqref{eq:fibpar}. Hence, we will disregard the term.

\subsection{Optical Schr\"odinger equation}
As the longitudinal spatial variations of the atomic densities $n_a$ are minute they do not cause reflections. Therefore, we can only consider forward-propagating wave trains
\begin{align}
\label{eq:Wellenvektorentkopplung}
\mathbf{E}_p(\vx, \omega)&= \mathbf{e}_p
 e^{ i k_p z } \mathcal{E}_p(\vx, \omega), 
\end{align}
with a carrier wave number  $k_p=\omega_p/c_0>0$ and amplitude 
$\mathcal{E}_p(\vx,  \omega)$ varying only slowly in the
$z$-direction ${\partial_z \mathcal{E}_p \ll k_p \mathcal{E}_p}$ and ${\partial_z^2 \mathcal{E}_p \ll k_p \partial_z \mathcal{E}_p}$.  
As we focus our discussion on optical frequencies,
i.\thinspace{}e. 
 $\omega\approx \omega_p\pm \Delta\omega$, we can also disregard the exponentially small term on the negative frequency side of Eq.~\eqref{Epfourier}.

For the simple effective three level system considered here, the specific polarization is not relevant. We therefore assume for simplicity linearly polarized laser fields. 
If the sublevels of the Zeeman manifolds  become relevant, effects like Faraday rotation can occur and a more sophisticated 
analysis is required.
In the absence of polarization-selective effects, we obtain the scalar optical paraxial Schr\"{o}dinger equation~\cite{Lax75,Marte97,Parax1}
\begin{align}
\label{eq:paraxSGL} 
 i \lslash_p \partial_z  
 \mathcal{E}_p(\vx, \omega)=  
 \left[-\tfrac{\lslash_p^2}{2 a_0^2} \Delta_\perp + 
 U(\vx,\omega)\right] \mathcal{E}_p,\\
\label{eq:Vton}
U(\vx, \omega) 
=\frac{\omega_p^2-\omega^2 \varepsilon_r(\vx,\omega)}{2\omega_p^2}
\approx - \tfrac{1}{2} \che(\vx, \omega).
\end{align}
Here, we have introduced a complex optical potential $U$ and approximated it by the susceptibility for pulses of frequency bandwidth  
$|\Delta\omega|\ll \omega_p\che_0$ (cf. App.~\ref{ap:Optpot}).
Due to the cylindrical geometry of the fiber with the hollow-core radius $a_0$, 
we introduce dimensionless polar coordinates $(r,\varphi)$ with 
$x=a_0 r \cos\varphi$ and $y=a_0 r \sin \varphi$ and a corresponding 
polar Laplacian operator $\Delta_\perp$.

The analogy to the quantum mechanical motion of a fictitious particle  moving in two dimensions with mass $a_0$ comes from identifying the smallness parameter of the theory 
$\lslash_p \leftrightarrow \hbar$ with the reduced Planck constant and the increasing propagation distance $z \leftrightarrow t$ with time. 
This is a typical result from the short-wave asymptotics of partial differential equations. However, the absorption in the complex optical potential renders the paraxial Schr\"odinger equation lossy. Thus, the Hamilton operator is not self-adjoint. It has  eigenvalues $\varepsilon(\delta)\in \mathds{C}$ and does not necessarily provide a complete set of orthogonal eigenmodes.

%
\subsection{Adiabatic propagation}
Its an experimental fact that the gas-filled hollow-core fiber
 supports well formed, time-delayed and attenuated probe pulses
  propagating downstream.  Therefore, we need to discuss the
   properties of the optical potential
   \begin{equation}
\label{eq:gaussian}
U(r,z;\delta) = - \frac{n_a(r,z) \alpha_0}{2}  f(\delta,\Delta,|\Omega_c(r)|).
\end{equation}
The spatial dependence of the Rabi frequency $|\Omega_c (r)|$ originates from the axially symmetric ground mode of the empty fiber
and exhibits a width $\sigma_d$. The magnitude of the control Rabi frequency is independent of the propagation distance $z$, as the control field couples the basically unpopulated states $\ket{2}$ and $\ket{3}$. 
In the considered experiments, the main radial and longitudinal dependence arises from the 
atomic density distribution $n_a(r,z)$, where the radial width of the atomic density distribution $\sigma_a < \sigma_d,a_0$ 
(cf. App.~\ref{ap:atomicdistr}).
This is in contrast to other reports \cite{HVM15, MSF95, MSF96}, 
where the inhomogeneity is dominated by $|\Omega_c(r)|$. 

According to 
Fig.~\ref{fig:system}, the HCF has short $l\ll L$, but still adiabatic 
$\lslash \ll l$,
entrance and exit 
zones \textcircled{1}, \textcircled{3}, where the density tapers off smoothly. This implies good mode matching 
$u_e^{(1)}(\text{\textcircled{1}}) 
\rightarrow u^{(1)}(\text{\textcircled{2}})
\rightarrow u^{(1)}_e(\text{\textcircled{3}})$
from the input to the exit port.
In general, this is a robust assumption for gaseous media, which do not exhibit sudden changes of the index of refraction. This assumption should also apply to experiments with HCFs either completely filled with room-temperature atoms \cite{BSV09,SMC14}, or partially filled with cold atoms.
Thus, we will study the eigenmodes of Eq.~\eqref{eq:paraxSGL} with an adiabatic factorization ansatz
\begin{align}
\label{eq:VollstLoesung}
 \mathcal{E}_p(r,\varphi,z;\delta) &=  
 e^{- i \phi(z)}
 u(r,\varphi,z; \delta)  \mathcal{E}_p(\delta) , \\
\phi(z)&=\int_0^{z} \text{d} \zeta \, 
q_p \varepsilon(\zeta;\delta),
\end{align}
 with $q_p=k_p/2 v^2$. 
Consequently, we have to find the eigenvalues and modes of the paraxial 
Schr\"odinger equation from 
\begin{equation}
\label{eq:stat}
  \varepsilon(z;\delta) u(r,z,\varphi; \delta) = 
  \left[-\Delta_\perp +  w(r,z; \delta)\right] u,
\end{equation}
with the rescaled optical potential 
\begin{align}
\label{eq:gaussianW}
w(r,z;\delta) =
 - w_0(r,z) f(\delta,\Delta,|\Omega_c(r)|),
\end{align}
where $w_0(r,z)=v^2 n_a(r,z) \alpha_0$.
We assume that the modes are well localized inside the fiber and pose the following hard boundary conditions
\begin{align}
\label{eq:cond}
 \partial_{r} u(r=0,\varphi,z;\delta) &= 0, &u(r=1,\varphi,z;\delta)=0,
\end{align}
valid  $\forall \varphi, z$. The modes are normalized  
$\int_{A}  \text{d}^2f \, |u|^2 = 1,$ over 
the fiber cross-section area $A$.

Given the eigenfunctions $u$ of Eq.~\eqref{eq:stat} are known, one can construct an energy functional according to 
\begin{equation}
\label{eq:redstat}
\varepsilon(z;\delta) = 
\int_{A} \text{d}^2f\,
 \left[ 
| \partial_r u|^2+\frac{|\partial_\varphi u|^2}{r^2}+
 w |u|^2  \right ].
\end{equation}
In analogy to wave-mechanics it consists of kinetic as well as potential energies.

Without an interferometric phase reference, we consider only the phase $\phi_2$ accrued in the long inner region \textcircled{2} $l\leq z \leq l+L$ and add the free phase $\eta=2l k_p(1- j_0^2/2 v^2)$  accumulated from zones \textcircled{1} and \textcircled{3} (cf. Eq.~\eqref{epshom}).
This results in an input-output relation at $z=2l+L$
\begin{gather}
\label{eq:r1}
\mathcal{E}^{(\text{out})}_p( \delta) = 
\mathcal{T}(\delta)
\mathcal{E}^{(\text{in})}_p( \delta), \\
\mathcal{T}(\delta)=e^{ i \eta+i k(\delta)L},\quad
k(\delta)=k_p - q_p \varepsilon(\delta).
\end{gather}
This transfer function $\mathcal{T}$ is the response of the inhomogeneously-filled HCF. For convenience, we have incorporated the carrier wave phase from Eq.~\eqref{eq:Wellenvektorentkopplung} into this definition as well.

\subsection{Stationary modes and spectrum}
The electrical eigenmodes and eigenvalues of the gas-filled HCF in zone \textcircled{2} are at the center of the pulse propagation problem. However, the 
problem is more involved as in the conventional theory of dielectric wave 
guides. There, one considers propagation in 
piece-wise constant  core and cladding material. Here, we have a smooth radially shaped density distribution, 
as well as strong EIT dispersion. 
This leads to
a focusing or defocusing effect for each spectral component, which is known as micro-lensing. In the following, we will analyze the eigenvalue problem for a homogeneously filled fiber and a two-layer model analytically, as well as the real Gaussian-shaped optical potential numerically.

Pulses with cylindrical symmetry are of most relevance for the experiment.
Therefore, we will compute axially symmetric modes $u(r;\delta)$ of
 the Schr\"odinger Eq.~\eqref{eq:stat}
\begin{equation}
\label{eq:neuMode}
\varepsilon u(r;\delta)= 
- \tfrac{1}{r}(r u')' +w(r;\delta)u .
\end{equation}
At first, it is interesting to first consider small local regions where the 
potential $w$ is almost constant. Then Eq.~\eqref{eq:neuMode} can 
be rephrased as a cylindrical Bessel differential equation \cite{Bessel} with index $\nu=0$
\begin{equation}~\label{eq:Bes}
\rho^2 h''(\rho) + \rho h'(\rho)+(\rho^2-\nu^2)h(\rho)= 0,
\end{equation} 
where $u(r)=h(\rho)$, $\rho=\kappa r$ and 
$\kappa^2 = \varepsilon-w \in \mathds{C}$.

Then, the solution of Eq.~\eqref{eq:neuMode}   reads
\begin{equation}
\label{eq:usol}
u(r) = a J_0(\kappa r) + b Y_0(\kappa r),
\end{equation}
which is a superposition of cylindrical Bessel functions of the first and second kind  $J_0$ and $Y_0$, respectively. It is important to note that $Y_0(r)$ diverges at the origin.

\subsubsection{Homogeneous potential}
A relevant special case is the 
radially homogeneously gas-filled fiber, where 
\begin{equation}
\label{constW}
w(\delta)= - w_0 f( \delta,\Delta,|\Omega_c|).
\end{equation}
Then, the general solution
Eq.~\eqref{eq:usol} has to satisfy the boundary conditions 
of Eq.~\eqref{eq:cond}. This leads to the discrete set of 
unnormalized eigensolutions 
\begin{align}
u^{(m)}(r; \delta) &= 
J_0(j_m r), \\
\label{epshom}
\varepsilon^{(m)}(\delta)&=j_{m}^2+ w(\delta) ,
\end{align}
where $\kappa=\{m\in \mathds{N}^{+},|
j_m=2.40, 5.52, 8.65, \ldots\}$ denote the $m$-th zero of $J_0(j_m)=0$. 
Thus, the eigenvalue $\varepsilon^{(m)}(\delta)$ disperses like  
$-\che(\delta)$ with a positive offset as shown in Fig.~\ref{fig:polariz}.

\subsubsection{Two-layer potential}
Partitioning the fiber radially at $r_1$ in two sections and defining 
a piece-wise constant optical potential as  
\begin{equation}
\label{eq:Vkasten}
  w^{(s)}(r, \delta) = 
  \begin{cases} w_1=w(r=0;\delta), & 0 \leq r < r_1 \\ 
  w_2=w(r_1;\delta), &r_1 \leq r < 1 
\end{cases},
\end{equation}
introduces a new degree of freedom into the system.
In particular when $w_2=0$ and $\Gamma_g=0$ \footnote{Any  potential offset can be removed with a
gauge transformation. In the new gauge, we can assume $w_2'=0$ and discuss $w_1'=w_1-w_2$.}, 
we can easily anticipate the response of the wave function $u$ to a changing two-photon detuning $\delta$.

\begin{figure}[t]
\includegraphics[scale=1]{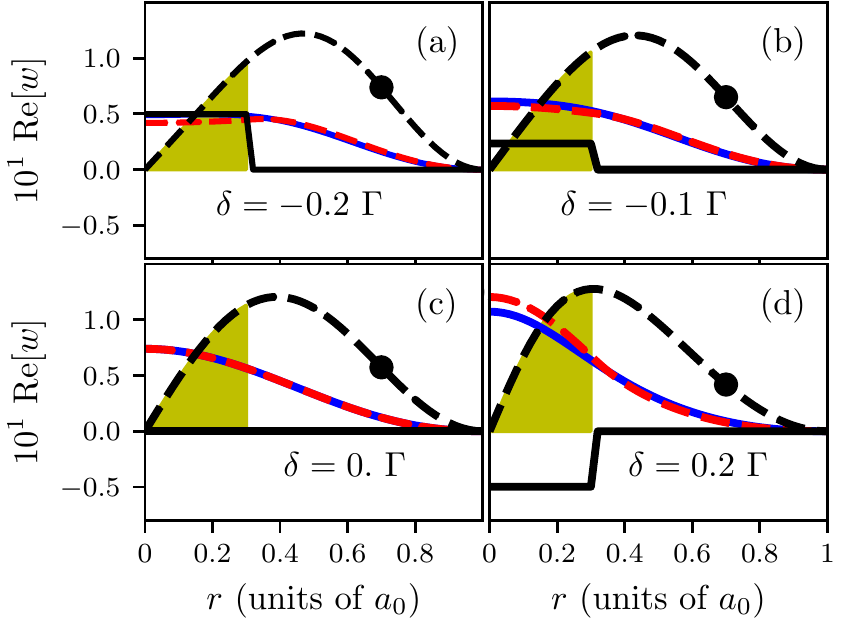}
\caption{
\label{fig:modes}
Real-part of the two-layer potential 
$\Re{w(r,\delta)}$ (solid black) versus radius $r$ for different 
two-photon detunings: (a) $\delta=\SI{-0.2}{\Gamma}$, 
(b) $\delta=\SI{-0.1}{\Gamma}$, (c) $\delta=\SI{0}{\Gamma}$, 
(d) $\delta=\SI{0.2}{ \Gamma}$ with other parameters $w_0 = 12.5$, $\Omega_c = \Gamma$, $\Delta=0$, $r_1 = 0.301$.
The corresponding radial intensities $|u^{(1)}|^2$ are shown for a Gaussian (blue, solid)
 and a two-layer potential (red, dashed) in arb. units.
The radially weighted intensity $r' |u^{(1)}|^2 $ 
(dashed-$\bullet$) are for the two-layer potential in arb. units. The yellow area is proportional to the potential energy of 
Eq.~\eqref{eq:redstat}.}
\end{figure}

From the susceptibility $\chi$ shown in Fig.~\ref{fig:polariz}, one can infer the behavior of the potential 
$w_1(\delta)\sim -\chi(\delta)$, as 
depicted in Fig.~\ref{fig:modes}. Thus, we infer that the dispersion 
$\chi'$ changes sign between the left and right zero crossings
located at 
$\delta^\pm$.
On the red side 
 of the resonance $\delta^-<\delta<0$,
the potential is repulsive $\Re{w_1(\delta)}>0$. This central hump enhances the spreading of the wave function like in a \textit{diverging lens}. 
On the blue side 
$0<\delta < \delta^+$, the potential has an attractive trough $\Re{ w_2(\delta)}< 0$, leading to light focusing similar to a \textit{converging lens}. This lensing effect is the basis of wave guiding in dielectric fibers with a piece-wise change in the index of refraction between the core and cladding material. However, in optical communication strongly frequency dependent materials are depreciated.  

The general solution for the modes of Eq.~\eqref{eq:usol} can be used to define the solutions in each of the two sections as
\begin{align}
\label{usectionn}
\begin{pmatrix}
u_{n}(r) \\
u'_n(r)
\end{pmatrix} 
&=M_n(r)
\begin{pmatrix}
a_{n} \\
b_{n}
\end{pmatrix},\\
M_n(r)&=
\begin{pmatrix}
J_0 (\kappa_n r) & Y_0 (\kappa_n r) \\
-J_1 (\kappa_n r) \kappa_n & -Y_1 (\kappa_n r) \kappa_n \\
\end{pmatrix},
\end{align}
with $\kappa_n^2=\varepsilon-w_n$ and $1\leq n\leq N=2$. Requiring that the solutions match smoothly in terms of value and gradient at the intersection $r_1$, one can define a  transfer matrix as
$T_n = M_{n+1}^{-1}(r_n)  M_{n}(r_n)$. With these defintions, the solution in the outer section reads
\begin{align}
\begin{pmatrix}
u_{2}(r) \\
u'_2(r)
\end{pmatrix} 
&=M_2(r)
T_1
\begin{pmatrix}
a_{1} \\
b_{1}
\end{pmatrix}.
\end{align}
The boundary conditions Eq.~\eqref{eq:cond} at the center of the fiber $u'(r=0)=0$ can be met by $a_1 = 1$, $b_1 = 0$. 
The required node 
of the mode at the outer boundary
\begin{equation}
\label{ueigenvalue}
u_{N}(r=1;\varepsilon(\delta))=0,
\end{equation}
defines a nonlinear equation for the complex eigenvalue 
$\varepsilon(\delta)$, which has to be found numerically for each value of the detuning $\delta$. 
For the simple two-layer model, Eq.~\eqref{ueigenvalue} can be expressed explicitly as $H(\kappa_1,\kappa_2,r_1)=0$ with
\begin{align}
\begin{aligned}
\label{eq:fullsqwenergy}
H=\frac{J_0(\kappa_1 r_1)J_1(\kappa_2 r_1)\kappa_2}{
J_0(\kappa_2 r_1)J_1(\kappa_1 r_1)\kappa_1}
-\frac{G_0(\kappa_2 r_1)
 - G_0(\kappa_2)}{
G_1(\kappa_2 r_1) - G_0(\kappa_2)},
\end{aligned}
\end{align}
introducing an auxiliary $G_i(x) \equiv Y_i(x)/J_i(x)$.

\begin{figure}[t]
\includegraphics[scale=1]{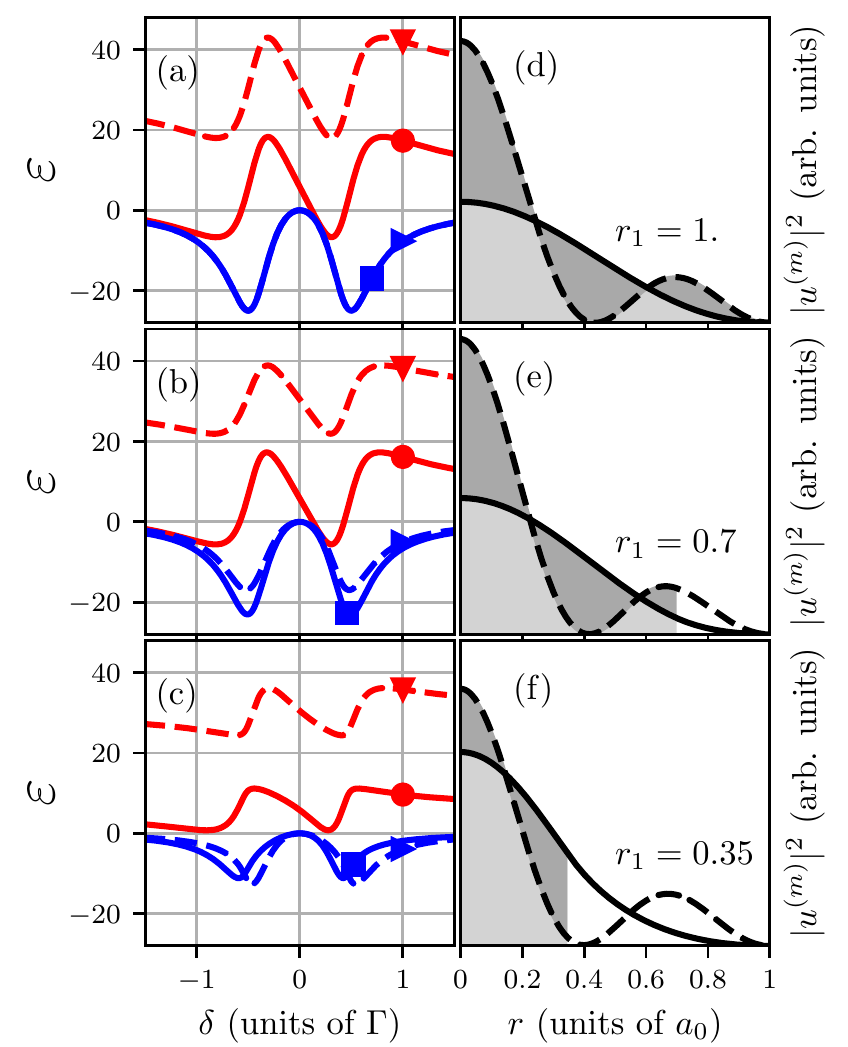}
\begin{flushleft}
\caption{
\label{fig:EVs1}
(a)-(c) Complex dispersion relation
$\varepsilon^{(1,2)}(\delta)$ 
 of the two-layer model versus two-photon detuning $\delta$: 
  $\Re{\varepsilon^{(1)}}$ ($\textcolor{red}{\bullet}$, solid), 
$\Re{\varepsilon^{(2)}}$ ($\textcolor{red}{\blacktriangledown}$, dashed), 
$\Im{\varepsilon^{(1)}}$ ($\textcolor{blue}{\blacksquare}$, solid), 
$\Im{\varepsilon^{(2)}}$ ($\textcolor{blue}{\blacktriangleright}$, dashed). 
(d)-(f) Corresponding intensities of eigenmodes 
$|u^{(1)}(r;\delta= \SI{0.2}{ \Gamma})|^2$ (black)  and  $|u^{(2)}(r;\delta= \SI{0.2}{ \Gamma})|^2$ (black, dashed) versus radius $r$ 
for different atomic distribution widths $r_1= \num{1} \text{(a;d)},0.7\,\text{(b;e)},0.35\,\text{(c;f)}$.
 The overlap between $|u^{(i)}|^2$  and the atomic medium is highlighted in gray colors. Note the different sign in the curvatures of $\Re{\varepsilon^{(m)}}$ for $r_1 = 0.35$ indicating a different sign of the micro-lensing effect. The other parameters are $(\Omega_c, \Delta, \Gamma_g) = (1, 0, 0)\, \Gamma$ and $w_0 =12.5$.}
\end{flushleft}
\end{figure}

In Fig.~\ref{fig:EVs1}, we depict the numerical solutions for
$\varepsilon^{(m)}(\delta,r_1)$ and the corresponding radial mode intensities for 
 $u^{(m)}(r,\delta,r_1)$ at $\delta = 0.2$. 
If the gas extends over the full width $r_1=1$, one recovers the result of the homogeneously filled fiber in Eq.~\eqref{epshom} [see Fig.~\ref{fig:EVs1}(d)].
Reducing the radius to $r_1 = 0.7$ reduces on the one hand the 
amplitude of the complex eigenvalues and on the other hand reveals 
an intermodal dispersion between $m=1$ and $2$ [see Fig.~\ref{fig:EVs1}(e)].
The shapes of the modes are altered only slightly indicating that 
lensing is very weak. 
For $r_1 = 0.35$ [see Fig.~\ref{fig:EVs1}(f)], a significant quadratic dispersion becomes noticeable due to micro-lensing. The effect has the opposite behavior for the two  modes considered.

\subsubsection{Arbitrarily-shaped potential}
In general, we need to solve the Schr\"odinger eigenvalue Eq.~\eqref{eq:stat} for a smoothly changing optical potential 
$w(r,\delta)$.
The transfer matrix method is a robust procedure to obtain the general solution for arbitrarily-shaped potentials. By partitioning  the integration domain into $N$ sections 
$0=r_0< r_1<\ldots <r_{N}=1$,
 one can assume that an optical potential  
\begin{equation}
w_n =  w(r_{n-1};\delta), \quad 1\leq n\leq N,
\end{equation}
is almost constant within each interval. 
Now the solution in each interval is given by Eq.~\eqref{usectionn}. Promoting the inner solution with the transfer matrix to the outer section, one finds 
\begin{align}
\begin{pmatrix}
u_{N}(r) \\
u'_N(r)
\end{pmatrix} 
&=M_N(r)
\prod_{n=1}^{N-1}
T_n
\begin{pmatrix}
1 \\
0
\end{pmatrix},
\end{align}
and Eq.~\eqref{ueigenvalue} defines the eigenvalues 
$\varepsilon^{(m)}(\delta)$ implicitly.
   
In Fig.~\ref{fig:modes}, we present the modes for 
the Gaussian potential  from Eq.~\eqref{eq:gaussian}
as well as the two-layer potential Eq.~\eqref{eq:Vkasten}. 
Correspondigly, Fig.~\ref{fig:compar} shows 
$\varepsilon^{(m)}$ containing the same number of atoms and 
having the same width 
$r_1=\sigma_a$. Both models agree well for detunings small compared to the EIT window width.  Deviations for larger detunings reflect the different large scale potential shape.  Yet, the two-layer model reproduces the relevant central dispersion with high accuracy.


\section{Pulse characterization}
\label{sec:lprop}

\subsection{Spectral output power}
\begin{figure}[h]
\subfigure{
\includegraphics[scale=1]{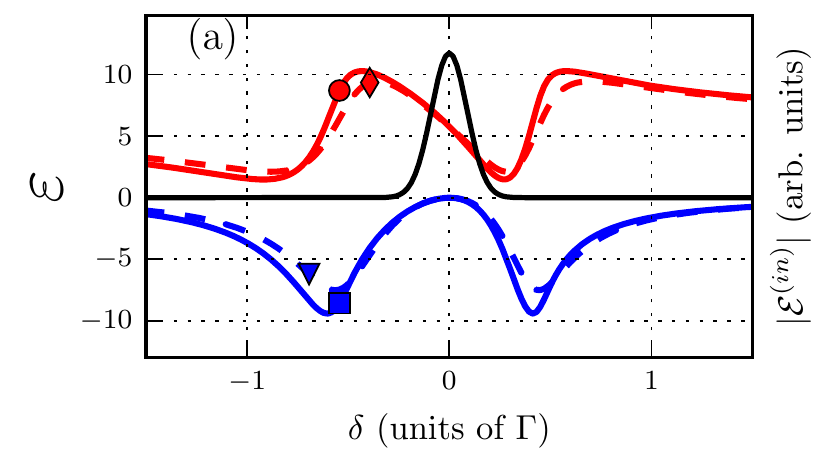}
\label{fig:pout}
}
\subfigure{
\includegraphics[scale=1]{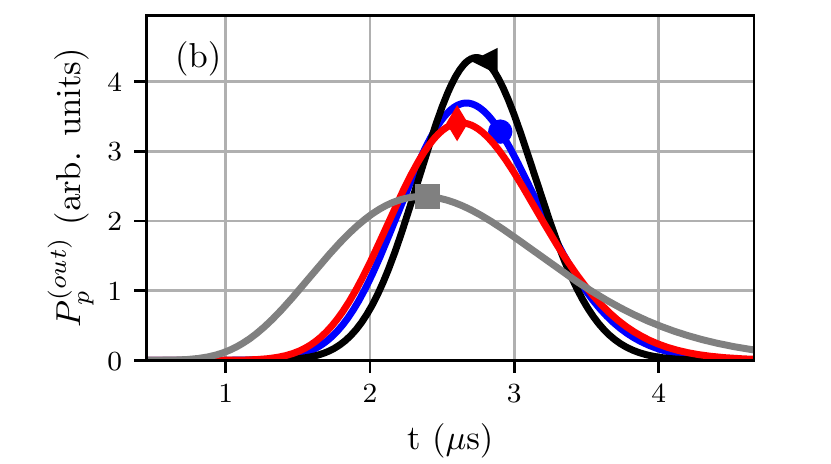}
\label{fig:pulse}
}
\begin{flushleft}
\caption{\label{fig:compar}
(a) Complex energies  versus two-photon detuning 
$\delta$: $\Re{\varepsilon^{(1)}}$ 
for two-layer 
(solid $\textcolor{red}{\bullet}$) and Gaussian 
(dashed $\textcolor{red}{\blacklozenge}$) potential 
as well as $\Im \varepsilon^{(1)}$ 
for both potential types 
(solid $\textcolor{blue}{\blacksquare}$, 
dashed $\textcolor{blue}{\blacktriangledown}$). For comparison, we superimposed the input amplitude 
$\mathcal{E}^{(in)}(\delta)$ (black, solid) in arbitrary units with
$w_0 = 12.5$, $\Omega_c = \Gamma$, $r_1 = 0.301$, $\Delta=0$ and other parameters as in Figs.~
\ref{fig:modes} and \ref{fig:EVs1}.
(b) Transmitted power 
$P_p^{(\te{out})}(t)$ in the two-layer potential
versus time $t$ for different  
one-photon detunings: $\Delta$,  
$\Delta/ \Gamma  = -1 \ (\textcolor{blue}{\bullet}),$ $ -0.5 \ (\textcolor{black}{\blacktriangleleft}), 0 \ (\textcolor{red}{\blacklozenge}), 0.5\ (\textcolor{gray}{\blacksquare})$ 
 with $r_1=0.301$, $w_0 = 25.0$, 
$\Omega_c = \Gamma$, $d_{opt} = 100, 
\tau_p =\SI{150}{\nano \second}$.}
\end{flushleft}
\end{figure}

A key feature of the complex transfer function $\mathcal{T}$ from
Eq.~\eqref{eq:r1} is the strong frequency dependence of the intensity transfer function $T(\delta)=|\mathcal{T}(\delta)|^2$ defined by
\begin{align}
\label{eq:attanuation}
I^\text{(out)}_p(\delta)&=T(\delta) I^\text{(in)}_p(\delta),
& T(\delta)=e^{-d_{\textrm{opt}}(\delta)},
\end{align}
which maps intensities $I=|\mathcal{E}|^2$. 
It follows from the imaginary part of the complex dispersion function and defines the optical density as
\begin{align}
\label{eq:opticaldensity}
d_{\textrm{opt}}(\delta) &=- 2 \theta\Im{ \varepsilon(\delta)}>0,
\end{align}
with  $\theta=q_p L$. Within the transparency window 
one can Taylor-expand [cf. Fig.~\ref{fig:pout}]
the complex dispersion as
\begin{equation}\label{eq:epsilonexpansion}
\varepsilon(\delta) = 
\epsilon + \epsilon'\delta + \frac{ \epsilon'' }{2}\delta^2 + \mathcal{O}(\delta^3),
\end{equation}
where $\epsilon=\varepsilon(0)$, $\epsilon'=\varepsilon'(0)$, etc. 
For vanishing ground state dephasing, 
the conventional definition of an optical density $d_{\textrm{opt}}(\delta=0)$ 
fails to be a good measure for light-matter interaction as it vanishes quadratically
\begin{align}
d_{\textrm{opt}}(\delta)&=\left(\frac{2 \delta}{\sigma_\text{EIT}}\right)^2+\ldots,  &\sigma_\text{EIT}=
\frac{2}{\sqrt{-\theta \Im{ \epsilon''}}}.
\end{align}
Here, we have defined a $1/e$ full-width $\sigma_{EIT}$ of a Gaussian distribution, from the imaginary part of the curvature of the dispersion relation.  

In analogy to the homogeneous EIT medium \cite{HHD99, FIM05,LFZ97},  
one obtains through this definition 
a more suitable optical density $d_\text{EIT}$ 
 and window width $\sigma_\text{EIT}$ 
\begin{align}
d_\text{EIT} &= 
\frac{|\Omega_c|^2 L}{\Gamma v_g },
&\sigma_\text{EIT} = 
\frac{|\Omega_c|^2}{\Gamma \sqrt{d_\text{EIT}}},
\label{eq:windowwidth}
\end{align}
also for an inhomogeneous HCF (see App.~\ref{shapeofdispersion}).
The inhomogeneous optical potential enters into this definition only through the modification of the group velocity 
\begin{align}
\label{eq:vspeed}
 v_g &=\frac{1}{\Re{k'(\delta=0)}}= - \frac{1}{q_p \Re{
 \varepsilon'(0)}}>0.
\end{align}
Please note that the unconventional signs arise from 
the negative exponent in the separation ansatz of
Eq.~\eqref{eq:VollstLoesung}.
If we use Eq.~\eqref{reepp} for the two-layer system without ground-state dephasing, we obtain
\begin{align}
\label{eq:vspeed2ls}
 v_g(r_1) &=\frac{1}{\mu(r_1)}
 \frac{|\Omega_c|^2 \lslash}{2 \Gamma n_a \alpha_0}.
\end{align}
This is the group velocity in a homogeneously filled HCF divided by a geometric factor, which looks approximately like 
$\mu\approx\sin{(\frac{\pi}{2}r_1)}^2$ 
as a function of the core radius $r_1$.

\subsection{Temporal output power}
Apart from the spectral response, we are also interested in the time-resolved output power 
$P_p(t)$ of the probe pulse emerging at the end of the fiber. Illuminating a photo detector with a field $\mathbf{E}_p(t)= 
\mathbf{E}_p^{(+)}(t)+\mathbf{E}^{(-)}_p(t)$,
that is formed by positive and negative frequency components 
causes an output signal \cite{cohentannoudjiBOOK1}
\begin{align}
P_p^\text{out}(t) = 2 c \varepsilon_0 \int_{A} \text{d}^2f \,
 \mathbf{E}^{(-)}_p(r,t)\cdot
 \mathbf{E}^{(+)}_p(r,t).
\end{align}
Thus, we have to Fourier-transform the output field in frequency space
\begin{align}
\label{Epfourierout}
\mathbf{E}_p^{(\text{out})}(r,\omega) &= 
\mathbf{e}_p  u^{(1)}_e(r)\left[
\mathcal{E}_p^{(\text{out})}(\omega)+
\mathcal{E}_p^{(\text{out})^\ast}(-\omega)\right] .
\end{align}
For the Gaussian probe pulse input amplitudes 
Eq.~\eqref{Epfourier} and the transmission function of Eq.~\eqref{eq:r1}, 
one obtains the real-time output amplitudes
\begin{align}
\label{eq:outpulse}
\mathcal{E}_p^{(\text{out})}(t) 
&=e^{-i[\omega_p t-k_p L-\eta]} \Phi(t) \mathcal{E}_p,\\
 \Phi(t)&=
\int_{-\infty}^\infty \text{d}\delta\,
\frac{ e^{-i \theta \varepsilon(\delta)-it \delta -\frac{\delta^2}{\Delta \omega^2}}}{2 \sqrt{\pi}\Delta \omega}.
\end{align}
Consequently, the observable output power reads
\begin{equation}
\label{outputpower}
P_p^{(\text{out})}(t) = 
2 c \epsilon_0 |\mathcal{E}_p^{(\text{out})}(t)|^2.
\end{equation}

In Fig.~\ref{fig:compar}, we  show the results of a numerical 
evaluation of the complex energies and the output pulse power 
Eq.~\eqref{outputpower} for a specific set of 
parameters and different one-photon detunings $\Delta$. 
The finite slope of the dispersion relation 
[see Fig.~\ref{fig:pout}], leads to a pulse delay. 
Moreover, the asymmetric dispersion translates into an asymmetric 
broadening and distortion of the output pulse for one-photon resonance 
($\Delta = 0$). However, the simulations in 
Fig.~\ref{fig:pulse} also show a pathway to minimize 
this broadening by walking off one-photon resonance with 
$\Delta \approx -0.5 \Gamma$.

\subsection{Time delay and distortion}
While the time delay of the probe pulse is a consequence of causal interactions with the atomic medium, the undesirable pulse distortion [see  Fig.~\ref{fig:pulse}] can be rectified to some degree. 
To be specific, we will assume the Gaussian input pulse is well localized within the EIT window. Then, we can use the second order expansion of dispersion relation Eq.~\eqref{eq:epsilonexpansion} and
evaluate the real-time transfer function as 
\begin{align}
\label{Phigauss}
 \Phi(t)&=
\frac{
e^{-i \theta \epsilon-
\frac{(t-\tau_\text{lag})^2}{\tau_p^2(1+i \xi)}
}}{2\sqrt{1+i\xi}},
&\xi=\frac{2 \theta }{\tau_p^2} \epsilon'' \in \mathds{C}.
\end{align}
Thus, the output pulse maintains its Gaussian shape with a 
positive time lag
\begin{align}
\tau_\text{lag}=-\theta \epsilon'=\frac{L}{v_g}>0,
\end{align} 
and pulse width $\tau^{\text{out}}_p>0$ that can be obtained from the complex denominator in the exponent of Eq.~\eqref{Phigauss}
\begin{equation}
\label{tauout}
\left(\tau^{\text{out}}_p\right)^2 \equiv 
\frac{(1- \Im \xi)^2 + (\Re{\xi})^2}{
1 - \Im\xi} \tau_p^2 \ge \tau_p^2,
\end{equation}
provided $\Im \epsilon''\leq 0$. Indeed, by inspection of the complex dispersion relation of  Fig.~\ref{fig:pout}, one finds a negative curvature of the  absorptive part. This is no coincidence but consequence of causality \cite{nussenzveig72}.

\subsection{Mitigation of lensing effects}~\label{sec:optpulse}
\begin{figure}[t]
\subfigure{
\includegraphics[scale=1]{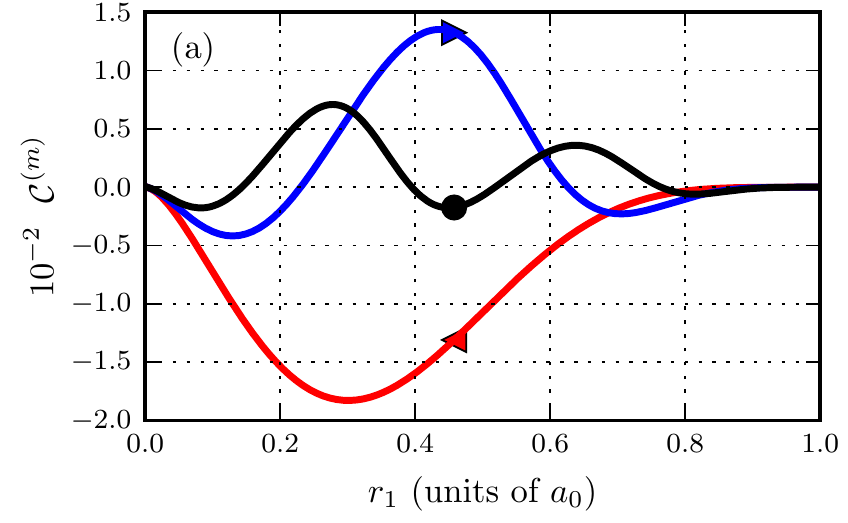}~\label{fig:refract}}
\subfigure{
\includegraphics[scale=1]{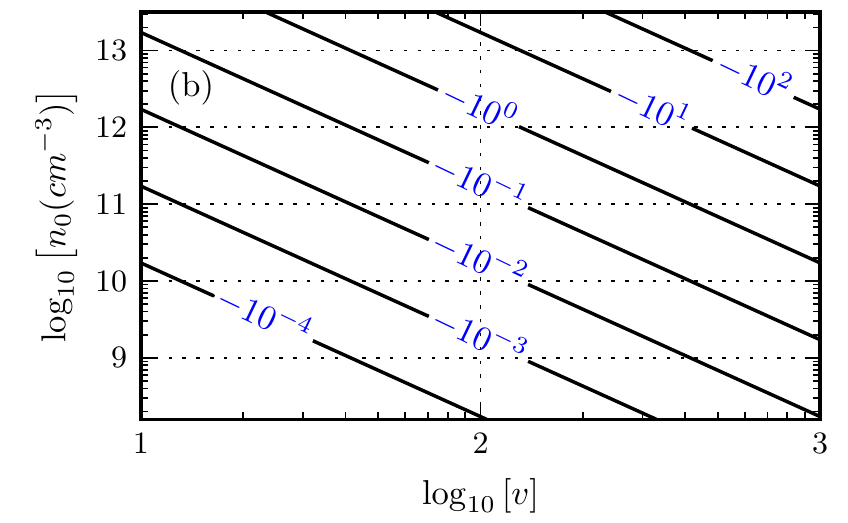}
\label{fig:groups}}
\caption{
(a) Refraction strength functions $\mathcal{C}^{(1,2,3)}(r_1)$ vs. width of gas filled core $r_1$. 
$\mathcal{C}^{(1)}$ shows the local minimum  $C^{(1)}(r_1=0.301) = -0.0183$ ($\textcolor{red}{\blacktriangleleft}$). $\mathcal{C}^{(2)}$ 
exhibits two zero crossings indicating a complete 
mitigation of lensing-effects ($\textcolor{blue}{\blacktriangleright}
$), while $\mathcal{C}^{(m)} $ for higher modes is an increasingly 
oscillating function with diminishing amplitude ($m=3$, $
\textcolor{black}{\bullet}$). 
(b) Contour plot of $\Delta_{\text{opt}}^{(1)}(v, n_0)$ for a specific experimental system using rubidium atoms ($^{87}$Rb $D_1$-line, $S_{FF'} = 1/6$) and $C^{(1)}(r_1 = 0.301)$.}
\end{figure}

The output pulse width points to a strategy to suppress the additional pulse broadening: $(\tau^{\text{out}}_p)^2 $ from Eq.~\eqref{tauout} is the sum 
of two widths that are added in quadrature. Therefore, the sum is minimal if 
\begin{equation}
\label{eq:optcond}
\Re{\varepsilon''(\delta=0,\Delta)} = 0.
\end{equation}
The shape of the output power is shown in
 Fig.~\ref{fig:pulse} and varies with the one-photon detuning $\Delta$. Using the explict expression for $\Re{ \epsilon''}$  from 
 Eq.~\eqref{reeppp}, 
 we can determine an optimal detuning as 
\begin{align}~\label{eq:detopt}
\Delta_{\textrm{opt}}^{(m)}(r_1)& = \Gamma w_0\  \mathcal{C}^{(m)}(r_1),\\
\textrm{with} \; \mathcal{C}^{(m)}(r_1)&=
\frac{H_{\varepsilon\varepsilon} H_{w}}{2 H_{\varepsilon}^2}
-\frac{H_{w\varepsilon}}{H_{\varepsilon}}
+\frac{H_{ww}}{2 H_{w}}.
\end{align}
The superscript $m$ has been restituted to account for 
different modes of the complex eigenenergy. We refrain from providing the explicit function $\mathcal{C}^{(m)}$, which is an uninspiring combination of Bessel functions exclusively depending on $r_1$. Instead we depict the shape functions 
$\mathcal{C}^{(1,2,3)}(r_1)$ in Fig.~\ref{fig:refract}.
While $\mathcal{C}^{(1)}(r_1)$ represents a strictly negative convex function with a local minimum at $r_1 = 0.301 $,
$\mathcal{C}^{(2)}$ has two zero crossings, indicating that a variation of $r_1$ for higher modes can lead to lensing with a different sign and even a complete mitigation of lensing for a certain $r_1$ [see Fig.~\ref{fig:EVs1}]. In Fig.~\ref{fig:groups}, we plot $\Delta_{\textrm{opt}}^{(1)}$ for different fiber parameters $v$ and atomic peak densities $n_0$ to show the expected magnitude.

In closing this section we note that this foregoing discussion relies on the transparency condition $w_1(0) = 0$. Ground-state dephasing with $\Gamma_g \neq 0$ breaks this condition and results in a nonlinear generalization of 
Eq.~\eqref{eq:detopt}.

\subsection{Suppression of micro-lensing}

Following the results of the last section, pulse-broadening induced by micro-lensing can be completely mitigated by applying a certain one-photon detuning $\Delta_{\textrm{opt}}^{(m)}$, that depends upon the parameters of the experimental setup [see Eq.~\eqref{eq:detopt}].
Using higher-order fiber modes ($m>1$) would allow for suppression of micro-lensing even at $\Delta=0$ for certain values $r_1$, as their corresponding functions $\mathcal{C}^{(m)}(r_1)$ have multiple zero crossings [see Fig.~\ref{fig:refract}]. However, these 
radially oscillating modes are experimentally undesirable.
Therefore, we discuss in the following only the suppression of 
micro-lensing for the fundamental mode. 
In case, one is restricted to using a certain one-photon detuning 
$\Delta$ and therefore cannot use this parameter for suppressing micro-lensing, $|\Delta_{\textrm{opt}}^{(1)} - \Delta|$ can be seen as an effective scale for the strength of the observed lensing effect. Thus, micro-lensing can be suppressed by minimizing 
$|\Delta_{\textrm{opt}}^{(1)} - \Delta|$ under a variation of other experimental parameters. For simplicity, we will set $\Delta = 0$ in the following.

In general, micro-lensing is reduced by choosing fibers with a rather small fiber parameter $v$ 
and for low atomic number density $n_a$, since $|\Delta_{\textrm{opt}}^{(1)}| \sim v^2 n_a$. 
The latter one, however, is in contrast to achieving a high light storage and retrieval efficiency, which depends linearly on the optical density $d_{\textrm{opt}}\sim n_a$. 
Therefore one might also think about a minimization of $|C^{(1)}(r_1)|$ to decrease the lensing strength. 
As Fig.~\ref{fig:refract} shows, this implies to either have $r_1 \rightarrow 1$ (homogeneous medium inside the fiber) or $r_1 \rightarrow 0$ (atoms concentrated near the fiber axis). 
The former approach leads to large collision rates of atoms with the fiber wall resulting in large decoherence. Therefore it is not a good solution.

The latter approach could be realized by modifying the atomic density distribution, e.g., via altering the ensemble temperature $T$ or the trapping potential depth $V_0$ (see App.~\ref{ap:atomicdistr}) as the atoms are inside the HCF. This will result in a change of the atomic number density $n_a=N_a/\pi r_1^2 L$ while leaving the number of atoms $N_a$ inside the HCF more or less constant. Therefore, we obtain
from Eq.~\eqref{eq:detopt}
\begin{align}
\Delta_{\te{opt}}^{(1)} (r_1)\biggr\vert_{N_a=const} &= 
\Gamma v^2 \alpha_0 \frac{N_{a}}{\pi L} \frac{\mathcal{C}^{(1)}(r_1)}{r_1^2}.
\end{align}
As $C^{(1)}(r_1)/r_1^2$ does not exhibit local extrema, but diverges for $r_1 \rightarrow 0$ \textit{in contrast} to $C^{(1)}(r_1)$, micro-lensing cannot be completely suppressed by a simply better localization of the atoms on the fiber axis. Nonetheless, we note that concentrating all atoms into a very small region $r_1 \rightarrow 0$, is an interesting limit. There, the optical potential approaches a two-dimensional Fermi pseudo potential $w(r) = -w_0 f(\delta) \delta^{(2)}(r) $ and many interesting analogies to condensed- or nuclear matter physics can be drawn \cite{zerorange88}.
We further note that the transverse localization will also be limited by the extension of the harmonic oscillator ground mode of the FORT potential. There, one might also have to consider that our theory is only valid for transverse extensions larger than the wavelength $\lambda_p$ (see Sec.~\ref{sec:parax}).

In order to avoid the divergence of $C^{(1)}(r_1)/r_1^2$ for $r_1 \rightarrow 0$, one would have to keep the atomic number density $n_a$ constant during the compression, i.e., reduce the number of atoms inside the HCF. The optimum detuning for constant density is then given by
\begin{align}
\Delta_{\te{opt}}^{(1)} (r_1)\biggr\vert_{n_a=const} &= 
\Gamma v^2 \alpha_0 n_a \mathcal{C}^{(1)}(r_1),
\end{align}
which vanishes for $r_1 \rightarrow 0$. Reducing the number of atoms loaded into the HCF, however, is of course contradicting the typical goal to maximize the optical depth for obtaining strong light-matter coupling and therefore usually better avoided.
This discussion illustrates, how the particularities of the experimental parameters critically influence mirco-lensing. Probably the best way to mitigate it is by choosing fibers of small core diameters, long lengths to keep the atomic number density as low as possible, and to concentrate the atoms near the fiber axis. If lensing then still occurs, it can be suppressed by an optimum one-photon detuning $\Delta$ according to Eq.~\eqref{eq:detopt}.

\subsection{Experimental observability}~\label{sec:comp}
In this chapter we describe how to observe the here discussed micro-lensing effects and discuss their estimated strength for several experimental systems. 
As a measure for the effect strength we use the magnitude of the one-photon detuning needed to suppress micro-lensing [see Eq.~\eqref{eq:detopt}] for the fundamental fiber mode.

\begin{figure}[t]
\includegraphics[scale=1]{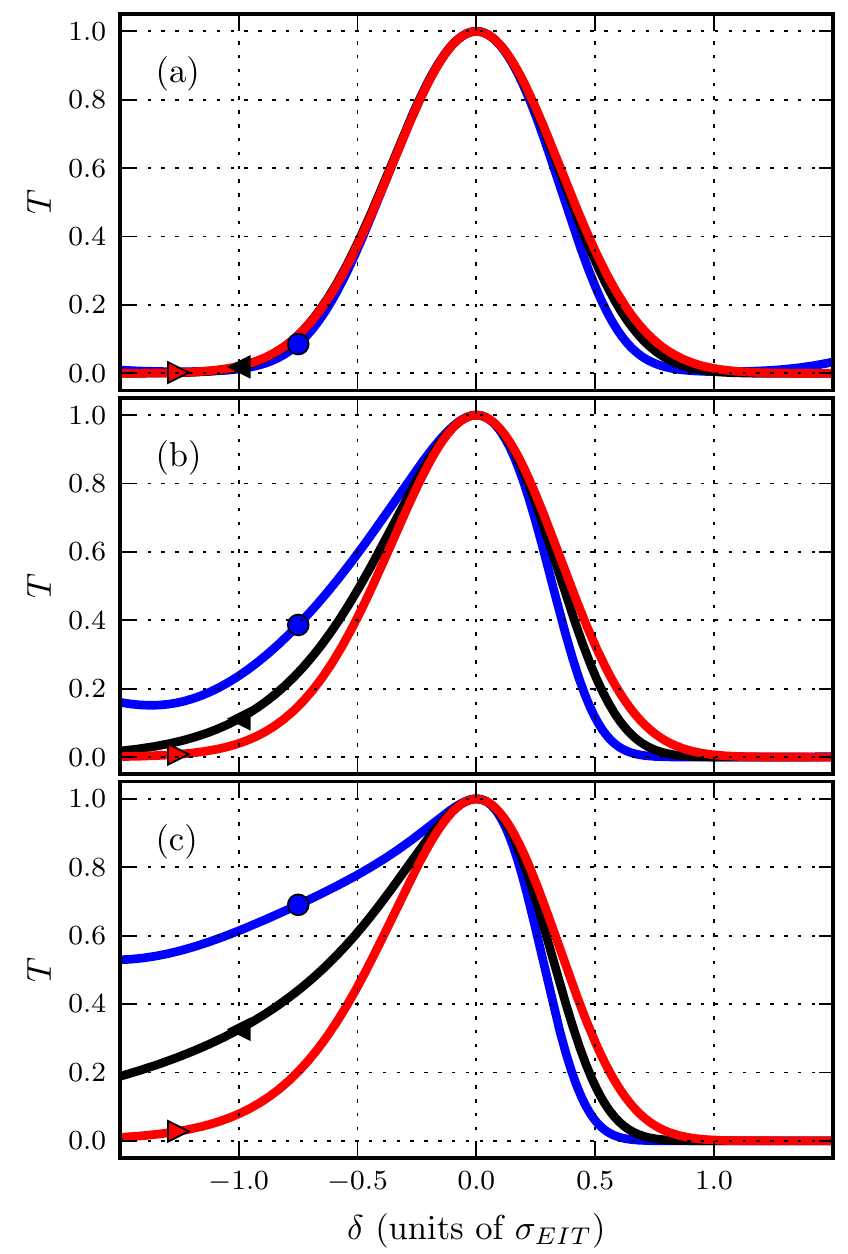}
\caption{
Intensity transmission $T(\delta)$ vs. detuning $\delta$ normalized to the EIT window width [see Eq.~\eqref{eq:windowwidth}]for different potential depths $w_0 = 5$ (a), $25$ (b), $50$ (c) with $r_1 = 0.301$ and different optical densities $d_{opt} = 5 \ (\textcolor{blue}{\bullet}), 25 \ (\textcolor{black}{\blacktriangleleft}), 250 \ (\textcolor{red}{\blacktriangleright})$. The one-photon detuning is $\Delta=0$. 
}
\label{fig:normalized_plots}
\end{figure}

The frequency-dependent micro-lensing will manifest itself in a more or less pronounced asymmetry in the fiber intensity transmission 
\begin{math}
T(\delta) = e^{ - d_{\textrm{opt}}(\delta)},
\end{math}
with optical density $d_{\textrm{opt}}(\delta)$ defined in 
Eqn.~\eqref{eq:opticaldensity}. This is shown in Fig.~\ref{fig:normalized_plots} for different potential depths $w_0$.
For the experimental setups using a small-core fiber with $a_0\sim3.5~
\mu$m \cite{BHB09,BHP14,BSH16,YB19pre} and a peak atomic density below 
$10^{12}~$cm$^{-1}$ \cite{BHP14,BSH16} lensing leads to minor 
modification of the transmission [Fig.~\ref{fig:normalized_plots}(a)]. 
Micro-lensing effects will therefore be currently hard to observe in 
such systems. For higher potential depths, i.e., medium-core fibers, 
micro-lensing effects should be easily observed. For instance, for 
$w_0 = 25$ and $w_0 = 50$  [Figs.~\ref{fig:normalized_plots}(b)\&(c)] 
defocusing for $\delta < 0$ leads to an enlarged transmission since 
the overlap of light and medium is decreased. For $\delta > 0$, on the 
other hand, the transmission window width will be reduced due to 
focusing. Asymmetries will be easier observed for low optical 
densities since in this case the transmission window is wider. In this 
case the lensing effect is not anymore dominated by $\Re{w}$. 
Instead $\Im w$ dominates and alters the lensing properties causing 
small deviations from the expected window widths. This can be seen in 
all three plots on the right edge of the transmission window.

As goes for the effect of micro-lensing on slow light, we plot in Fig.~\ref{fig:groups} the required optimal control beam detuning
$\Delta^{(1)}$ as a function of the fiber parameter $v$ [see Eqn.~(\ref{eq:fibpar})], which is proportional to the ratio of fiber core radius and probe wavelength, and the peak atomic number density $n_0$ [see Eqn.~(\ref{eq:density})]. As can be seen, the larger 
the core radius, the lower the atomic density required to obtain a 
significant detuning $\Delta^{(m)}\neq0$. If, e.g., we consider again the experimental setup using a small-core fiber \cite{BHP14,BSH16}, the optimum detuning will be in the range of $\Delta_{\textrm{opt}}^{(1)}\sim0.03\Gamma$ and therefore micro-lensing effects will be 
currently hard to observe with such a setup. On the other hand, for the medium-core fibers with $v\sim 200$ the same atomic density will lead to $\Delta_{\textrm{opt}}^{(1)}\sim 1\Gamma$ and therefore micro-lensing (and its suppression) will be more relevant in these systems.

\section{Conclusions}~\label{sec:sum}

We developed a simple two-layer  model for light propagation in an 
atom-filled HCF under EIT conditions to derive an effective description of micro-lensing effects. 
We showed that a highly dispersive interior of a linear wave guide acts like a confined lens with a highly dispersive curvature causing frequency-dependent focusing and defocusing inside the fiber. 
Since we assumed perfect adiabatic guidance this lensing manifests itself in an effectively distorted intramodal light dispersion, modifying group-velocity and absorption due to modulation of the light-matter coupling. 
The magnitude and sign of this lensing effect depends critically on 
the distribution width of the dispersive medium and on the field 
distribution of the propagating mode (intermodal dispersion), i.e., 
fiber core diameter.

We further proved in the framework of the developed two-layer model, that lensing-induced intramodal dispersion can be compensated in the center of the EIT window by applying an optimal 
one-photon detuning. This detuning has a non-linear dependence on the 
distribution width of the atomic medium and also serves as an 
effective reference for the lensing strength. With the help of this 
reference we showed that current experimental realizations using 
medium-core HCFs are expected to show micro-lensing effects, whereas 
small-core HCFs will be subject to micro-lensing for larger densities 
only.

\begin{acknowledgments}
R.W. acknowledges support for travel from the German
Aeronautics and Space Administration (DLR) through Grant
No. 50WM 1557. 
\end{acknowledgments}
%
\appendix
\section{Thermal atomic density distribution}
\label{ap:atomicdistr}
Atoms are loaded into a HCF by use of an auxiliary far off-resonant red-detuned laser beam in the ground mode $u^{(1)}_e(r)$ which establishes an optical dipole potential \cite{GWO00} 
\begin{equation}
V_\text{dip}(r) = - V_0 |u^{(1)}_e(r)|^2,
\end{equation}
of width $\sigma_d$  and sufficient depth $V_0>0$
to localize the high field seeking atoms in the center of the fiber. After collisional relaxation, a thermal atomic density distribution emerges as
\begin{align}
n_a(r) &= \tilde{n}_0 e^{ - \frac{V_\text{dip}(r)}{k_B T}}, 
&N_a=\int \text{d}^3x \,n_a
\end{align}
at temperature $T$ with particle number $N_a$. If the temperature is low enough,  
$k_b T \ll V_0$, one can Taylor-expand the dipole potential around the center to 
obtain a generic Gaussian atomic density distribution 
\begin{equation}~\label{eq:density}
n_a(r) = n_{0} 
e^{- \frac{r^2}{\sigma_a^2}},
\end{equation}
with a thermal width
$ \sigma_a(T) = \sigma_d \sqrt{k_B T / V_0}$.

 \section{Complex optical potential}~\label{ap:Optpot}
In the slowly varying envelope approximation, we assume that the amplitude 
$\mathcal{E}(\vx,\omega)$ has a finite support only in a small frequency domain around the carrier frequency $\omega_p$ of the probe pulse. 
By introducing a parameter $\zeta=(\omega-\omega_p)/\omega_p$, this
condition reads $|\zeta|\ll 1$ and the optical potential reads 
\begin{align}
U&= \frac{\omega_p^2 - \omega^2 \varepsilon_r}{2 \omega_p^2} = - \frac{\che}{2}  
- \varepsilon_r (\zeta 
- \frac{1}{2}\zeta^2).
\end{align}
Now, one can easily see that the first term dominates the expression by considering the magnitude of the ratio of $\che$ and the second term
\begin{equation}
\frac{|\che|}{|2 \varepsilon_r \zeta|} \approx 
\frac{\che_0 \omega_p}{\Omega_c} \approx \frac{\che_0 \omega_p}{\Gamma} \approx 10^{6}.
\end{equation}
Here, we have assumed that the pulse bandwidth matches the EIT window
$|\omega-\omega_p|=\Omega_c = \Gamma=\SI{10}{\mega\hertz}$, 
$\omega_p = \SI{1}{\peta\hertz}$ and $\che_0 = 10^{-2}$.\\

\section{Raman transition in the interaction picture \label{App:IAP}}
Given the Raman configuration of Fig.~\ref{fig:CS}
 and coupling of the three atomic levels by a monochromatic control laser
$\vec{E}_c(t)=\Re{[\vec{e}_c\mathcal{E}_ce^{-i \omega_c t}]}$
and a probe laser
$\vec{E}_p(t)=\Re{[\vec{e}_p\mathcal{E}_p e^{-i \omega_p t}]}$,
the dipole interaction energy reads 
$\hat V(t)=-\hat{\vec{d}}\cdot  (\vec{E}_p(t)+\vec{E}_c(t))$. 
In the optical domain, the rotating-wave approximation 
\citep{cohentannoudjiBOOK1}
applies and the
atomic energy Hamilton operator reads
\begin{align}
 \hat{H}&(t)/\hbar = 
 \omega_1  \hat\sigma_{11}+
 \omega_2  \hat\sigma_{22}+ 
 \omega_3  \hat\sigma_{33} \\
&+\tfrac{1}{2}\left(\hat\sigma_{31}\Omega_p e^{-i \omega_pt}
 + 
 \hat\sigma_{32} \Omega_c e^{-i \omega_c t}
 +\text{h.c.}\right).\notag
\end{align}
Here, $\hbar \omega_i$ is the energy of the state $\ket{i}$, 
the transition operators are defined as 
$\hat\sigma_{ij} = \ket{i} \bra{j}$, $\vec{d}_{ij}$ are dipole matrix elements and the Rabi frequencies 
$\Omega_p=-\vec{d}_{31}\vec{e}_p\mathcal{E}_p/\hbar$ and $\Omega_c=-\vec{d}_{32}\vec{e}_c\mathcal{E}_c/\hbar$ quantify the coupling strength of the corresponding transition. 
In order to eliminate oscillating amplitudes in the Schr\"{o}dinger picture, one transforms to an interaction picture 
$\ket{\psi(t)} = \hat{U}(t) \ket{\psi(t)}'$
with
\begin{align}
\hat{U}(t) =&\exp{[-i(\omega_3-\omega_p \hat\sigma_{11}-\omega_c  \hat\sigma_{22})t]}.
\end{align}
This results in a interaction picture Hamilton operator
\begin{align}
\hat{H}'/\hbar =& \hat{U}^\dagger(t)\left[  
\hat{H}(t)/\hbar-i\partial_t\right]  \hat{U}(t) \\
 =& 
 (\Delta+\delta)\hat\sigma_{11}+
 \Delta \hat\sigma_{22}+\tfrac{1}{2}\left( \hat\sigma_{31} \Omega_p
 + 
 \hat\sigma_{32}\Omega_c
 +\text{h.c.}\right),\notag
\end{align}
where $\Delta=\omega_c-\omega_{32}$ denotes the one-photon detuning and $\delta=\omega_p-\omega_c-\omega_{21}$ the two-photon detuning as in Eq.~\eqref{eq:HCS}.
In a matrix representation this reads
\begin{align}
H'_{ij}=
\hbar\begin{pmatrix}
\delta+\Delta & 0&\frac{\Omega_p^\ast}{2}\\
0 & \Delta &\frac{\Omega_c^\ast}{2}\\
\frac{\Omega_p}{2} & \frac{\Omega_c}{2}&0
\end{pmatrix}.
\end{align}

\section{Shape of the complex dispersion}
\label{shapeofdispersion}
In the considered parameter range, the analytical two-layer model and the Gaussian potential model predict similar dispersions close to resonance, as shown in  Fig.~\ref{fig:pout}. Therefore, we will pursue the two-layer model, assuming that there are no atoms in the outer fiber ($w_2 = 0$) and disregard ground-state  dephasing $\Gamma_g = 0$. This implies $w_1(\delta=0) = 0$ on resonance.

With the aid of the implicit function theorem, 
we can determine all coefficients  
$\partial^n_\delta\varepsilon(0)$ of the Taylor series 
of the dispersion relation Eq.~\eqref{eq:epsilonexpansion} 
from  
Eq.~\eqref{eq:fullsqwenergy} 
\begin{align}
H(w(\delta),\varepsilon(\delta))=0,
\end{align}
which defines a relation between the complex energy $\varepsilon$ and $w\equiv w_1$ as functions of $\delta$ implicitly.
On resonance, we 
have a solution $\mathcal{P}=(w(0),\varepsilon(0)) = (0,j_1^2)$ where  the medium is transparent.
If $\mathcal{P}$ is a regular point, we can obtain higher derivatives from the condition
\begin{equation}~\label{eq:deltan}
\partial_\delta^n H (w(\delta), \varepsilon(\delta)) = 0.
\end{equation}
For the first and second derivatives of the complex dispersion, we find 
\begin{align}
\label{eq:e1}
\epsilon' &=\mu
  w',\quad \mu=- \frac{H_{w}}{ H_{\varepsilon}},\\
 \label{eq:e2}
\epsilon'' &= \mu w''
- \frac{
\epsilon'^2 H_{\varepsilon\varepsilon}  + 
2  \epsilon' w' H_{w\varepsilon}
+ w'{}^{2} H_{w w}}{H_{\varepsilon}},  
\end{align} 
where we have abbreviated partial derivatives of $H$ with subscripts and denote $w'=w'(0)$ and $w''=w''(0)$.  The partial derivatives can be evaluated explicit in terms of Bessel functions, i.\,e.
\begin{align}
\mu&=\frac{\pi  r_1^2 j_{1} Y_0\left(j_{1}\right) \left(J_0\left(
r_1  j_{1}\right){}^2+J_1\left(r_1 
   j_{1}\right){}^2\right)}{2 J_1\left(j_{1}\right)}.
   \end{align}
Higher order expressions can be calculated but are not shown.

We can also obtain explicit values for the potential derivatives on resonance from Eqs.~\eqref{fvanishgammag} and \eqref{eq:gaussianW}
as
\begin{align}
w'&= -\frac{w_0}{\delta_\text{EIT}} , 
&w'' = 
\frac{2 \Delta + i \Gamma}{\Gamma \delta_\text{EIT}} w'.
\end{align} 

Due to the fact that all partial derivative of $H$ evaluated at the stationary point are real, one can explicitly evaluate the Taylor coefficients of the dispersion series as
\begin{align}
\Re{\epsilon} &=j_1^2,\quad \Im \epsilon=0,\\
\label{reepp}
\Re{\epsilon'}&=\mu\Re{w'}, \quad \Im{\epsilon'}=0,\\
\label{reeppp}
\Re{\epsilon''}&=\mu \Re{ w''}-
 \frac{
\epsilon'^2 H_{\varepsilon\varepsilon}  + 
2  \epsilon' w' H_{w\varepsilon}
+ w'{}^{2} H_{w w}}{H_{\varepsilon}},\\
\Im{\epsilon''}&=\mu \Im{w''}.
\end{align}

\bibliographystyle{FEIT_final.bbl}

\end{document}